\documentclass{emulateapj}
\usepackage{apjfonts,graphicx}

\newcommand{\Msun}{\ensuremath{M_\odot}}

\newcommand{\ergcms}{\ensuremath{\mathrm{erg~cm}^{-2}~\mathrm{s}^{-1}}}
\newcommand{\chandra}{\textit{Chandra}}
\newcommand{\cxo}{\textit{Chandra X-ray Observatory}}

\newcommand{\hubble}{\textit{Hubble Space Telescope}}
\newcommand{\xmm}{{\it XMM}}

\newcommand{\Fx}{\mbox{$F_{\rm x}$}}
\newcommand{\kms}{\mbox{km s$^{-1}$}}

\newcommand{\pg}{PG\,1115+080}

\begin{document}
\shorttitle{Microlensing of \pg}
\shortauthors{Pooley et al.}
\slugcomment{submitted to ApJ}

\title{The Dark-Matter Fraction in the Elliptical Galaxy Lensing the Quasar \pg}

\author{D.\ Pooley\altaffilmark{1}, S.\ Rappaport\altaffilmark{2},  J.\ Blackburne \altaffilmark{2}, P.\ L.\ Schechter\altaffilmark{2}, J.\ Schwab\altaffilmark{2}, J. Wambsganss\altaffilmark{3}}  

\altaffiltext{1}{Astronomy Department, University of Wisconsin--Madison, 475 North Charter Street, Madison, WI  53706, {\tt dave@astro.wisc.edu}}
\altaffiltext{2}{Department of Physics and Kavli Institute for Astrophysics and Space Research, MIT, Cambridge, MA 02139}
\altaffiltext{3}{Astronomisches Rechen-Institut, Zentrum f\"ur Astronomie der Universit\"at Heidelberg, Moenchhofstr.12-14, 69120 Heidelberg, Germany}


\begin{abstract}
We determine the most likely dark-matter fraction in the elliptical galaxy quadruply lensing the quasar \pg\ based on analyses of the X-ray fluxes of the individual images in 2000 and 2008.  Between the two epochs, the $A_2$ image of \pg\ brightened relative to the other images by a factor of six in X-rays.  We argue that the $A_2$ image had been highly demagnified in 2000 by stellar microlensing in the intervening galaxy and has recently crossed a caustic, thereby creating a new pair of micro-images and brightening in the process. Over the same period, the $A_2$ image has brightened by a factor of only 1.2 in the optical.  The most likely ratio of smooth material (dark matter) to clumpy material (stars) in the lensing galaxy to explain the observations is $\sim$90\% of the matter in a smooth dark-matter component and $\sim$10\% in stars.  
\end{abstract}

\section{Introduction}

The theory of gravitational lensing is by now quite well understood \citep[e.g.,\,the review by][]{1999fsu..conf..360N}.  For the case of a quasar quadruply imaged by an intervening galaxy, a very simple model for the lensing potentials  --- a monopole plus a quadrupole --- usually succeeds in fitting the positions of quasar images at the 1--2\% level.  However, it has become increasingly clear that these same models do considerably worse at fitting the {\em relative fluxes} from quasar images \citep[e.g.,][]{2004ApJ...610...69K,2002ApJ...567L...5M}. Such ``flux ratio anomalies'' are thought to be the product of small scale structure in the gravitational potentials of the lensing galaxies.  

There are two leading explanations for this small scale structure.  One intriguing explanation is that we are seeing \textit{milli}-lensing by dark matter condensations of sub-galactic mass \citep{1995ApJ...443...18W,1998MNRAS.295..587M,2002ApJ...572...25D, 2001ApJ...563....9M, 2002ApJ...565...17C}, which are predicted in large numbers in $N$-body simulations. However, the much more likely explanation (and exciting for very different reasons) is that the anomalies are largely the result of \textit{micro}-lensing by stars in the intervening galaxy \citep{1995ApJ...443...18W, 2002ApJ...580..685S}.

If the flux ratio anomalies are due to {\em milli}-lensing, i.e., $10^4$\,--\,$10^8$~\Msun\ dark matter condensations \citep{1992ApJ...397L...1W}, then the Einstein radii of such masses, projected back to the quasar, are sufficiently large that we would expect the flux ratios to (i) be the same at all wavelengths, and (ii) remain constant with time (except for source variability).

In fact, neither of these expectations based on {\em milli}-lensing is observed, and the results are overwhelmingly more compatible with stellar {\em micro}-lensing.  If this is indeed the case, then (i) the stellar Einstein radii projected back to the quasar are more nearly comparable in size with the expected quasar emission regions, thereby allowing for a probe of the inner regions of their accretion disks; (ii) variations in the flux ratio anomalies with time in one source, or from source to source, can provide a direct measure of the dark-to-stellar matter ratios at projected radial distances of $\sim$$2-6$ kpc in elliptical galaxies; and (iii) the flux ratio anomalies are expected to vary dramatically on time scales as short as a few years.  We note in passing that item (i) above works because microlensing, in effect, is the most powerful zoom lens in astronomy, probing angular sizes down to $\sim$$10^{-6}$~arcsec, which is a factor of $\sim$100 better than even VLBI.

Recently we have systematically analyzed ten quadruply-imaged quasars using \cxo\ archival data and \hubble\ visible images \citep{2007ApJ...661...19P}. We find that the flux ratio anomalies in the X-ray images of quads are systematically larger than for the same quads imaged in the visible, by a factor of $\sim$2.  As expected from the models \citep[see][]{2002ApJ...580..685S, 2007ApJ...661...19P}, it is the highly magnified saddle-point image among the four images that is most susceptible to stellar microlensing.  \citet{2007ApJ...661...19P} concluded that the extent of the quasar accretion disks in the optical (i.e., $r_{\rm opt}$) must be comparable with (i.e., $\gtrsim 1/3$) the Einstein radii, $r_{\rm ein}$, of the stellar microlenses in order to reduce the flux ratio anomalies by the factor of $\sim$2 observed.  This conflicts with values of $r_{\rm opt}$ calculated from simple accretion-disk models, which are expected to be considerably smaller than the Einstein radii, with typical ratios of $r_{\rm opt}/r_{\rm ein}$ in the range of only $0.01 - 0.3$, with a median value of $0.04$ \citep{2007ApJ...661...19P}.  Thus, the observationally inferred optical emission regions in quasars, based on microlensing, are much larger than anticipated.  This is an intriguing mystery to be pursued.

\begin{figure}[t]
\centering
\includegraphics[width=0.235\textwidth,height=0.235\textwidth]{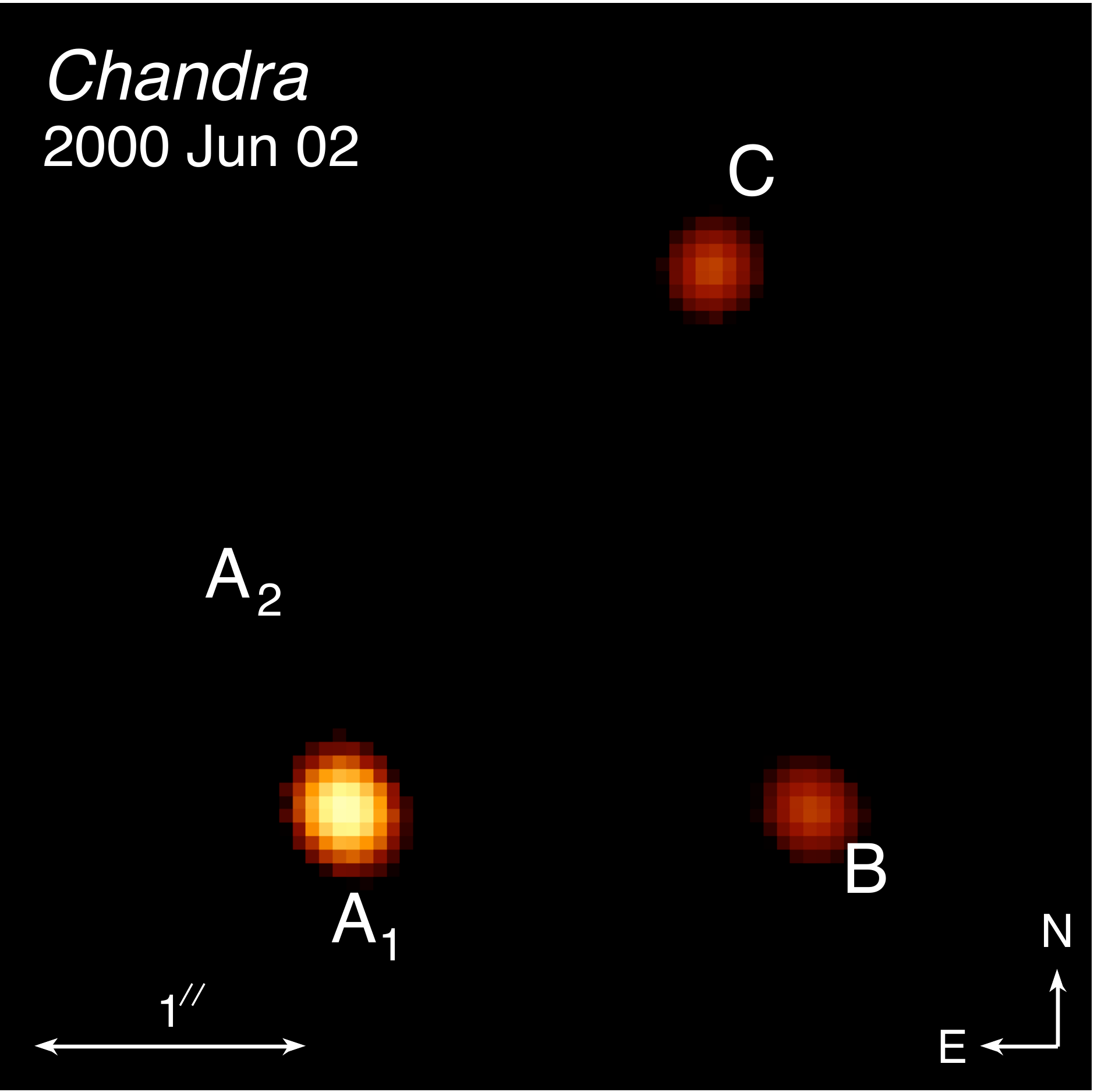}\hglue0.015\textwidth                      
\includegraphics[width=0.235\textwidth,height=0.235\textwidth]{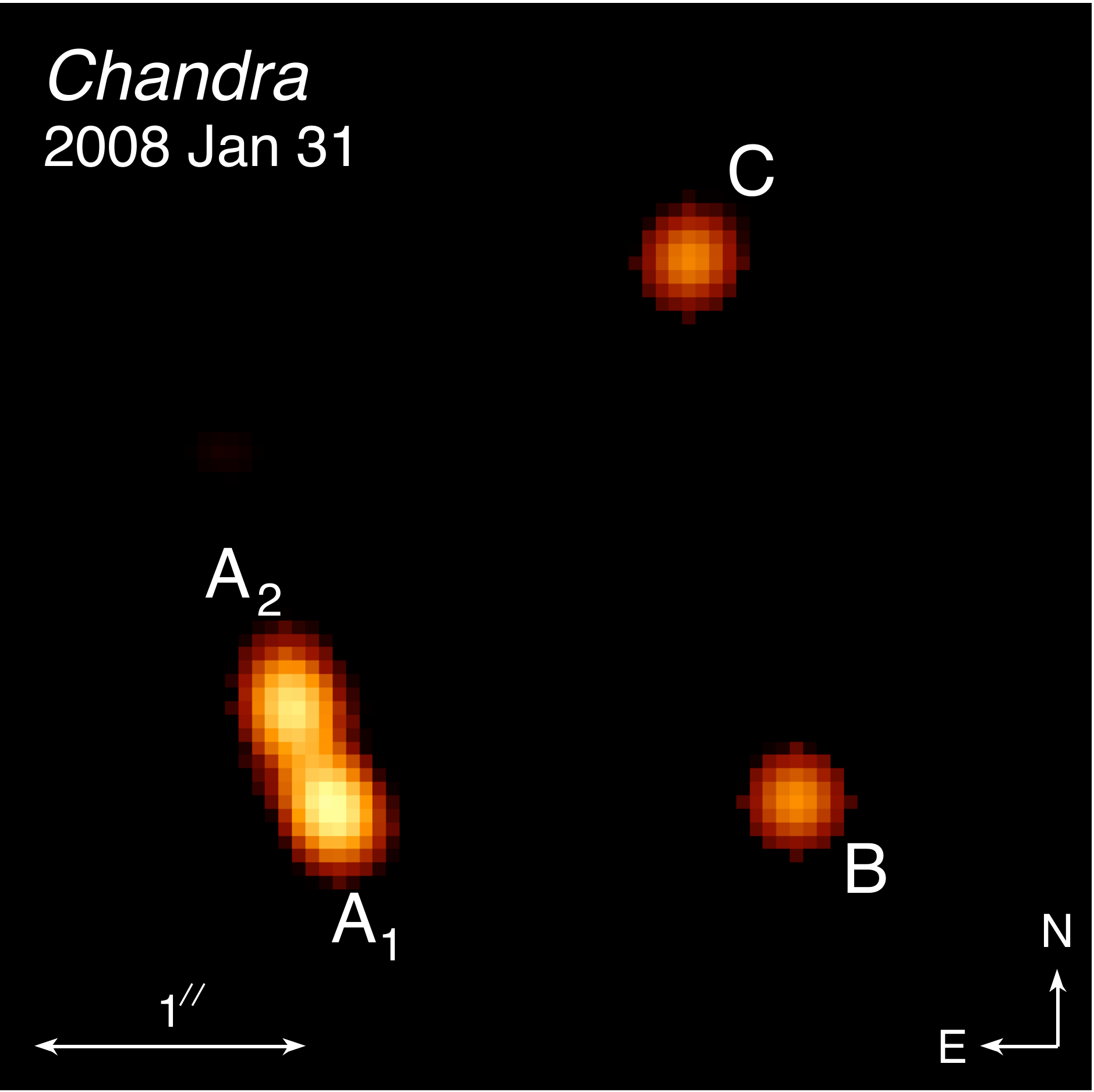}\vglue0.015\textwidth                                                                                       
\includegraphics[width=0.235\textwidth,height=0.235\textwidth]{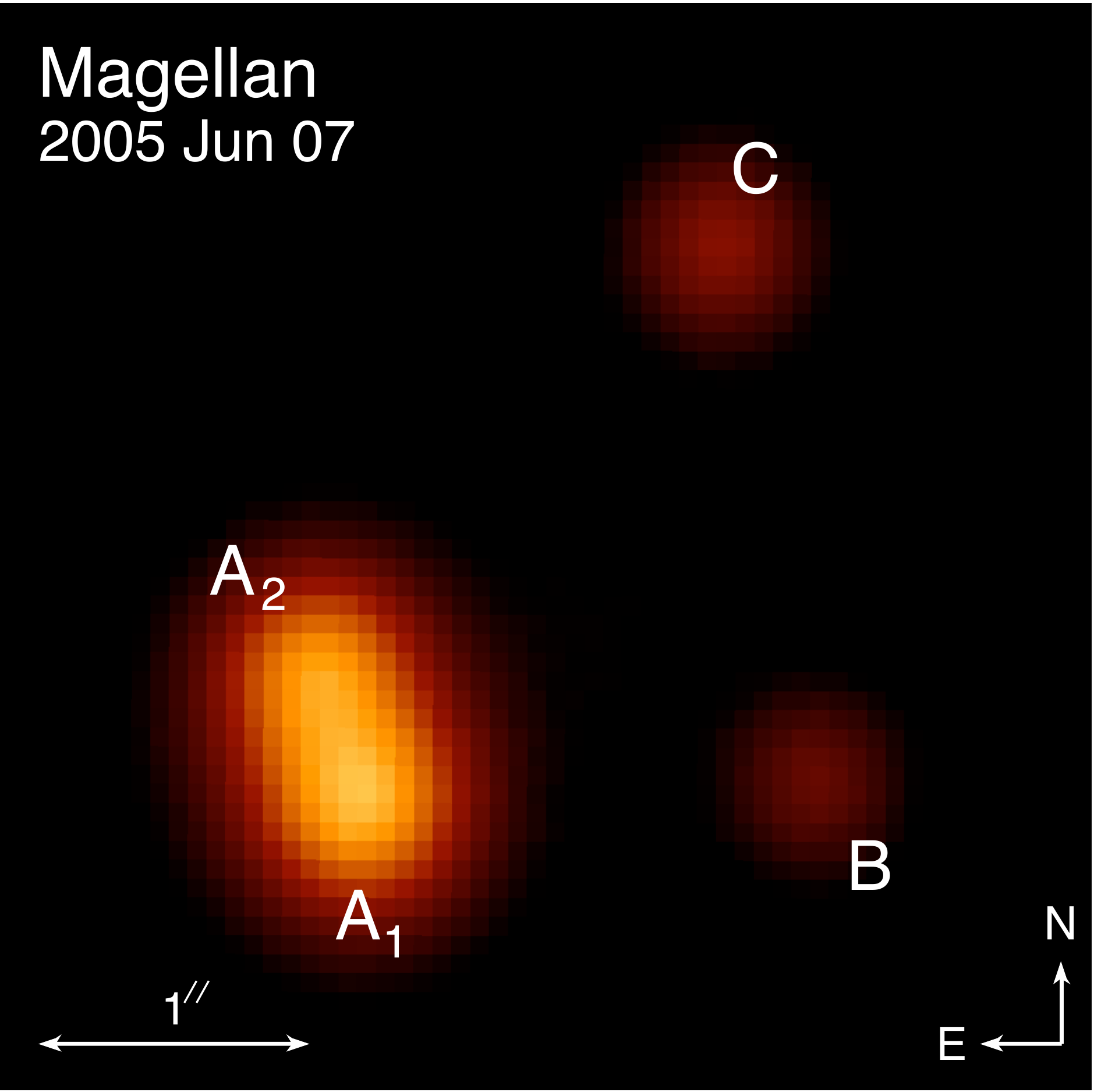}\hglue0.015\textwidth                       
\includegraphics[width=0.235\textwidth,height=0.235\textwidth]{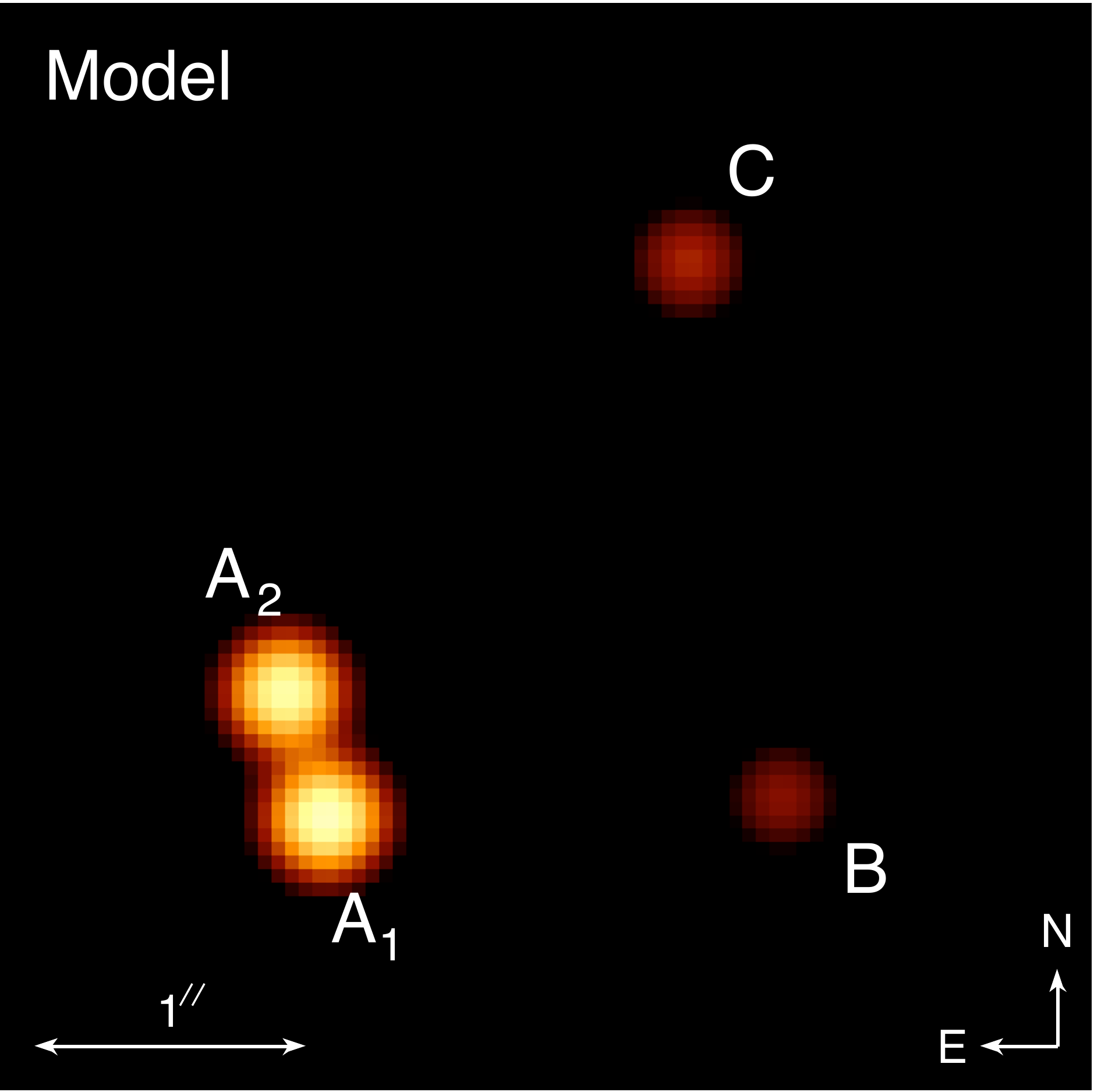}                                 
\caption{Images of \pg.  Clockwise from upper left: Maximum-likelihood reconstructed \chandra\ image from 2000, smoothed by a Gaussian; maximum-likelihood reconstructed \chandra\ image from 2008, smoothed by a Gaussian; expected image predicted from singular isothermal sphere plus shear model of lensing galaxy; Magellan $i'$-band image. Each image is $4''\times4''$.  The apparent lack of any X-ray emission from $A_2$ in 2000 is an artifact of the maximum-likelihood reconstruction.}
\label{fig:images}
\end{figure}

Of equal astrophysical significance, the same observations of the amplitudes and frequency of occurrence of X-ray flux ratio anomalies can also be used to infer the fraction of dark matter at distances from the center of the lensing elliptical galaxies corresponding to the impact parameter of the images \citep[typically $\sim$2--6 kpc;][]{2004IAUS..220..103S}. In this paper we pursue this latter line of investigation for the quad lens \pg.  In particular, we describe a new \chandra\ observation of \pg\ in January 2008 which indicates that the $A_2$ image has dramatically brightened in X-rays compared to its state in 2000 (see Fig.\,\ref{fig:images}).  In \S2.1 we review prior optical and X-ray observations of \pg, while in \S2.2 we present the new \chandra\ observations and describe the analysis by which we determined the flux ratios.  In \S3 we describe how inferences about the dark-to-stellar matter ratio can be made from observations of microlensing.  Finally, in \S4 we summarize our results.

\section{Observations of \pg}

\subsection{Prior Observations of \pg}

\pg\ was the second gravitationally lensed quasar to be discovered \citep{1980Natur.285..641W} and the first one found to be quadruple.  It has been the subject of numerous studies at wavelengths ranging from radio to mid-infrared to optical to UV to X-ray.  It was the first gravitational lens to yield multiple time delays \citep{1997ApJ...475L..85S}, and it shows uncorrelated variations among its images \citep{1985A&A...149L..13F}.  The brightest pair of images, $A_1$ and $A_2$, is quite close ($\sim$$0.5''$), and simple lens models have these two images resulting from a ``fold'' caustic.  In such cases \citep{KGP05} one expects the two images to be very nearly equal in brightness.  

\begin{figure}
\centering
\includegraphics[width=0.49\textwidth]{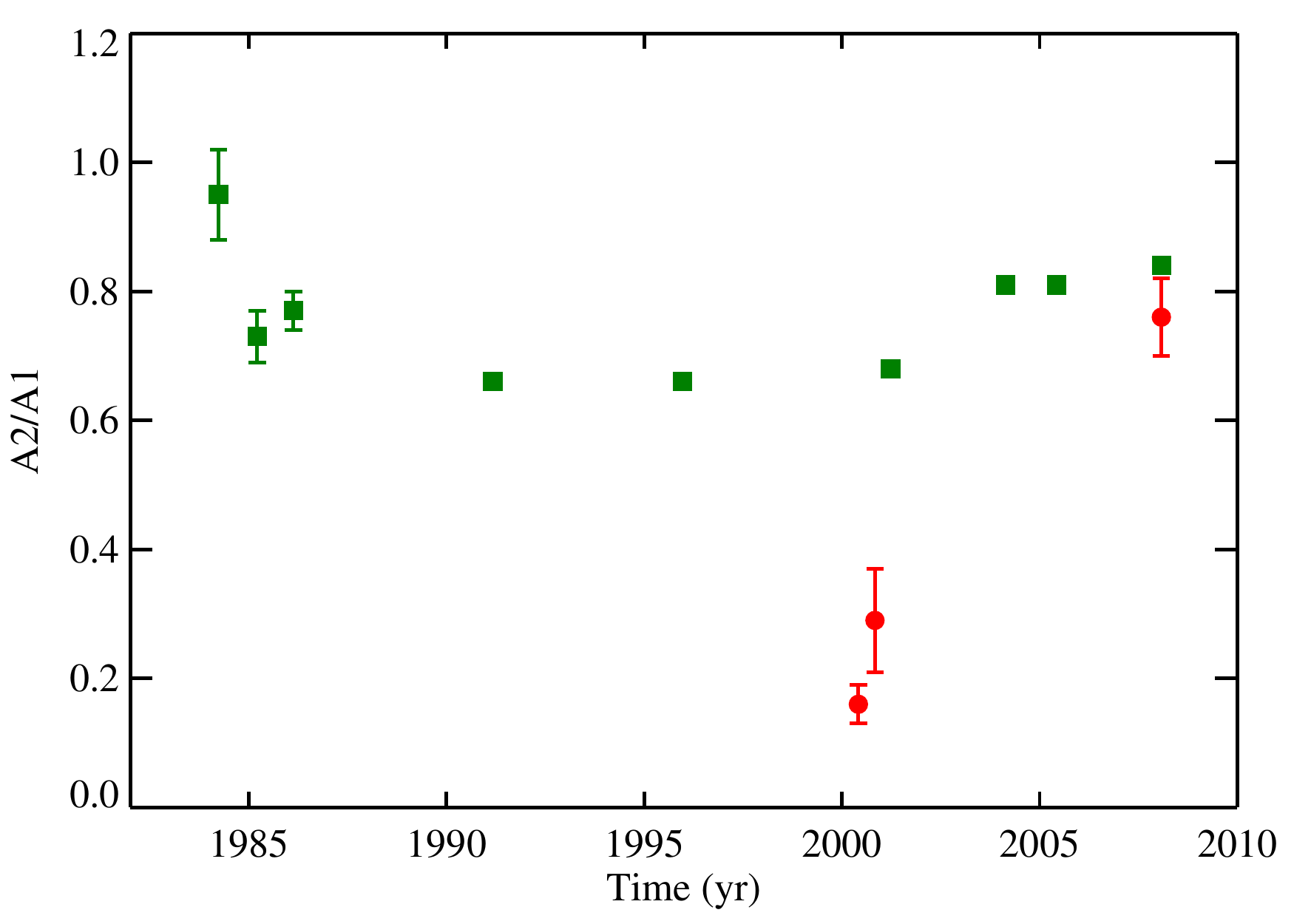}
\caption{Long-term history of the flux ratio $A_2/A_1$ of \pg\, in the optical band (green squares) and in the X-ray band (red circles). For some of the observations, the plotted error bars are smaller than the plotting symbols.}
\label{fig:ltlc}
\end{figure}

From its discovery more than a quarter century ago to the present, the optical flux ratio between images $A_2$ and $A_1$ has been in the range of $\sim$0.65--0.85 as determined from numerous measurements (see Fig.\,\ref{fig:ltlc}), and \citet{1986A&A...158L...5V} reported $A_2/A_1$ varying by $\sim$30\% on a timescale of one year via measurements taken on electronographic plates.  \citet{2005ApJ...627...53C} found that $A_2/A_1$ is nearly unity in the mid-IR.  The nominal flux ratio from the lens {\em model} is $A_2/A_1 = 0.96\pm 0.05$.  Thus, the {\em optical} flux ratio anomaly is slight, and nearly constant in time (see Fig.\,\ref{fig:ltlc}).  By sharp contrast, the first two \chandra\ observations in 2000 yielded X-ray ratios of $A_2/A_1 = 0.16\pm 0.03$ and $0.29 \pm 0.08$, with the $A_2$ component dramatically demagnified \citep{2006ApJ...648...67P}.  This extreme anomaly is very similar to the case of SDSS\,0924$+$0219 in the optical \citep{2006ApJ...639....1K}.  The arrangement of the images is virtually identical in the two systems.  Later \xmm\ observations of \pg\ (which could not resolve the individual quasar images) showed an overall increase in the X-ray flux of the system, which \citet{2006ApJ...648...67P} speculated could be due to a brightening of $A_2$.  This became the motivation for undertaking the \chandra\ observation reported here.

\subsection{2008 \chandra\ observation of \pg}

\pg\ was observed for 28.8 ks on 2008 January 31 (ObsID 7757) with the Advanced CCD Imaging Spectrometer (ACIS). The data were taken in timed-exposure mode with an integration time of 3.24 s per frame, and the telescope aim point was on the back-side illuminated S3 chip. The data were telemetered to the ground in very faint mode.

Reduction was performed using the CIAO 4.0 software provided by the \chandra\ X-ray Center. The data were reprocessed using the CALDB 3.4.3 set of calibration files (gain maps, quantum efficiency, quantum efficiency uniformity, effective area) including a new bad pixel list made with the acis\_run\_hotpix tool. The reprocessing was done without including the pixel randomization that is added during standard processing. This omission slightly improves the point-spread function. The data were filtered using standard event grades and excluding both bad pixels and software-flagged cosmic-ray events. No intervals of strong background flaring were found.

Our analysis follows the procedure laid out in \citet{2007ApJ...661...19P}. We produced a 0.3--8 keV image of \pg\ with a resolution of 0.0246$''$ per pixel.  To determine the intensities of each lensed quasar image, a two-dimensional model consisting of four Gaussian components plus a constant background was fit to the data.  The background component was fixed to a value determined from a source-free region near the lens.  The relative positions of the Gaussian components were fixed to the separations determined from \hubble\ observations \citep{1993AJ....106.1330K}, but the absolute position was allowed to vary.  Each Gaussian was constrained to have the same full-width at half-maximum, but this value was allowed to float.  The fit was performed with Sherpa 3.4 using Cash (1979) statistics and the Powell minimization method.  From this fit, we measure the value of $A_2/A_1$ to be $0.76\pm0.06$, very near to the optical flux ratio.

In order to visualize the dramatic rise in the flux of $A_2$ (with the $A_2$ and $A_1$ images clearly separated), we produced maximum likelihood reconstructions of two \chandra\ images from the 2000 and 2008 observations.  For this, we used the max\_likelihood function in the IDL Astronomy User's Library, which is based on the algorithms of \citet{1972JOSA...62...55R} and \citet{1974AJ.....79..745L}.  This is a simple, iterative, Bayesian technique to estimate the deconvolution of the observed data and the instrumental point spread function (PSF).  The PSF was constructed using the \chandra\ Ray Tracer (ChaRT) to produce a simulated PSF and Marx 4.3\footnote{\url{http://space.mit.edu/ASC/MARX/}} to project the PSF onto the detector.  \chandra's PSF is energy dependent, and we extracted the spectrum of \pg\ to provide the appropriate input to ChaRT.  With this simulated PSF of \pg, we performed 1000 iterations of the max\_likelihood function on the data and smoothed the result with a Gaussian for aesthetic reasons.  The results are shown in Fig.~\ref{fig:images}, along with a Magellan $i'$-band image and an image using four Gaussians to represent the expected image based on a singular isothermal sphere plus shear model of the lensing galaxy.  The apparent lack of any emission from $A_2$ in 2000 is an artifact of the Lucy-Richardson deconvolution, which does not seem to robustly handle faint sources (e.g., $A_2$) in the immediate vicinity of much brighter sources (e.g., $A_1$).

\begin{figure}
\centering
\includegraphics[width=0.45\textwidth]{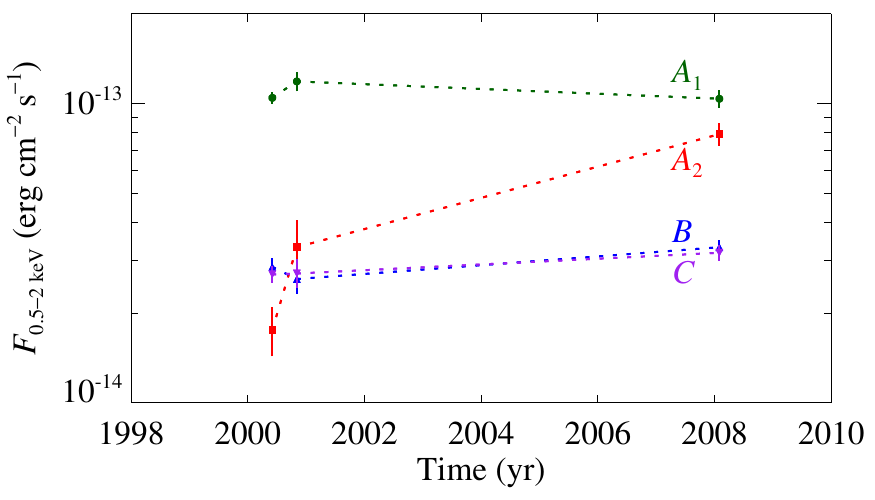}
\caption{X-ray fluxes vs. time of the individual images of \pg.}
\label{fig:xlc}
\end{figure}

The long-term history of the $A_2/A_1$ flux ratio in the optical and X-ray is summarized in Fig.\,\ref{fig:ltlc}.  The X-ray history of the fluxes from each of the quasar images, based on our two-dimensional Gaussian fits and the measured spectrum of image $C$, is given in Table~\ref{tab:xrayflux} and displayed in Fig.~\ref{fig:xlc}, showing that the change in $A_2/A_1$ is indeed a result of $A_2$ becoming less demagnified.

\begin{deluxetable}{lllll}
\tablewidth{0pt}
\tablecaption{X-ray fluxes of \pg\ images}
\tablehead{\colhead{Date} & \multicolumn{4}{c}{\Fx\ ($10^{-14}$ \ergcms)} \\
\colhead{} & \colhead{$A_1$ (HM)} & \colhead{$A_2$ (HS)} & \colhead{$B$ (LS)} & \colhead{$C$ (LM)}}
\startdata
2000 Jun 02& $10.5\pm0.5$ & $1.75\pm0.33$ & $2.85\pm0.19$ & $2.69\pm0.18$\\
2000 Nov 03& $11.9\pm0.9$& $3.32\pm0.73$ & $2.60\pm0.29$ & $2.71\pm0.30$\\
2008 Jan 31& $10.4\pm0.7$ & $7.92\pm0.68$ & $3.32\pm0.20$ & $3.18\pm0.19$
\enddata
\tablecomments{HM = high-magnification minimum image; HS = high-magnification saddle-point image; LS = low-magnfication saddle-point image; LM = low-magnfication minimum image}
\label{tab:xrayflux}
\end{deluxetable}

\begin{figure}
\centering
\includegraphics[width=0.45\textwidth]{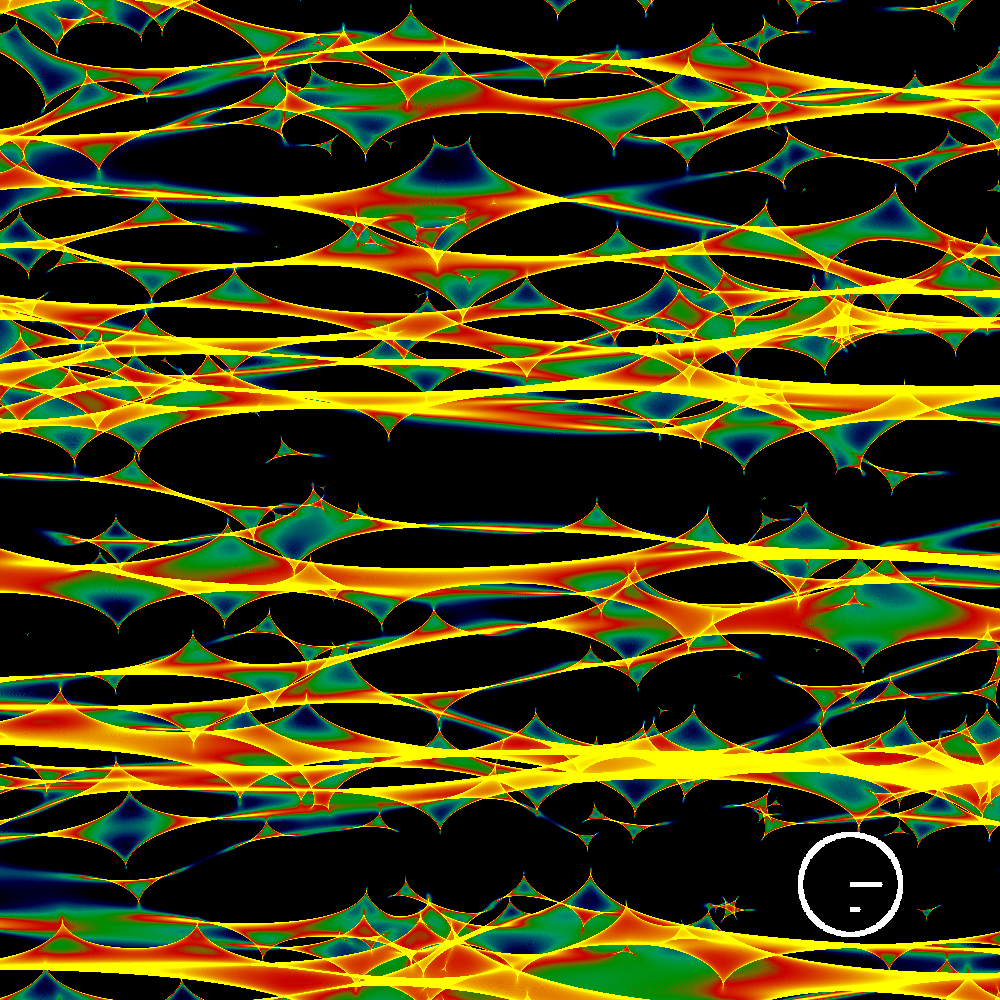}
\caption{Sample magnification map for an overall lensing potential which includes 80\% dark matter and a 20\% contribution from stars (for a negative parity image region with a mean magnification of 12).  This map represents a tiny section of the lensing galaxy that is $\sim$ 50 microarcseconds on a side, and is presented with a logarithmic display with the log of the {\em mean magnification} of the galaxy subtracted off.  For example, yellow regions correspond to magnifications of greater than a factor of $\sim$3; dark regions are demagnifications of greater than a factor of $\sim$3.  The white circle has a radius equal to the Einstein radius of a stellar microlens.  The white bars indicate how far bulk motion would typically shift the overall pattern in 8 years for transverse velocities of 300 and 1000 \kms; the direction is arbitrary.}
\label{fig:Wambs}
\end{figure}

\section{Evaluation of Dark-to-Stellar Matter Content}

The way in which observations of flux ratio anomalies, and their variations with time, can lead to an estimate of the dark-to-stellar matter content of the lensing elliptical galaxy is based on analyses of stellar microlensing magnification maps \citep{1999JCoAM.109..353W}.  The magnification distributions change with the addition of smooth matter; they get more asymmetric and in particular allow for larger demagnifications than for only stellar lenses \citep[e.g.][]{2002ApJ...580..685S}.   An illustrative magnification map for a lensing galaxy with 80\% smoothly distributed dark matter and 20\% stars is shown in Fig.~\ref{fig:Wambs}.  This type of magnification map is constructed in the source plane, and its center is referenced to the location of one of the quad images.  It shows the effects of microlensing magnification (due to the sum of all the microimages) for a source location anywhere within the map.  For the visual presentation of the map, the mean magnification, due to the smooth lensing potential, has been subtracted off.  The particular example shown in Fig.\,\ref{fig:Wambs} is for a highly magnified saddle (HS) point image (e.g., $A_2$ in \pg).  The sharp-edged features in Fig.\,\ref{fig:Wambs} correspond to caustics, the crossing of which by the source corresponds to the creation or annihilation of microlensing image pairs.  Such magnification maps have also been generated so as to additionally represent the high-magnification minimum (HM), low-magnification saddle (LS), and low-magnification minimum (LM) images ($A1$, $B$, and $C$, respectively).

We approach the magnification map analyses in three slightly different and complementary ways, described and discussed below.

\begin{figure}[t]
\centering
\includegraphics[width=0.49\textwidth]{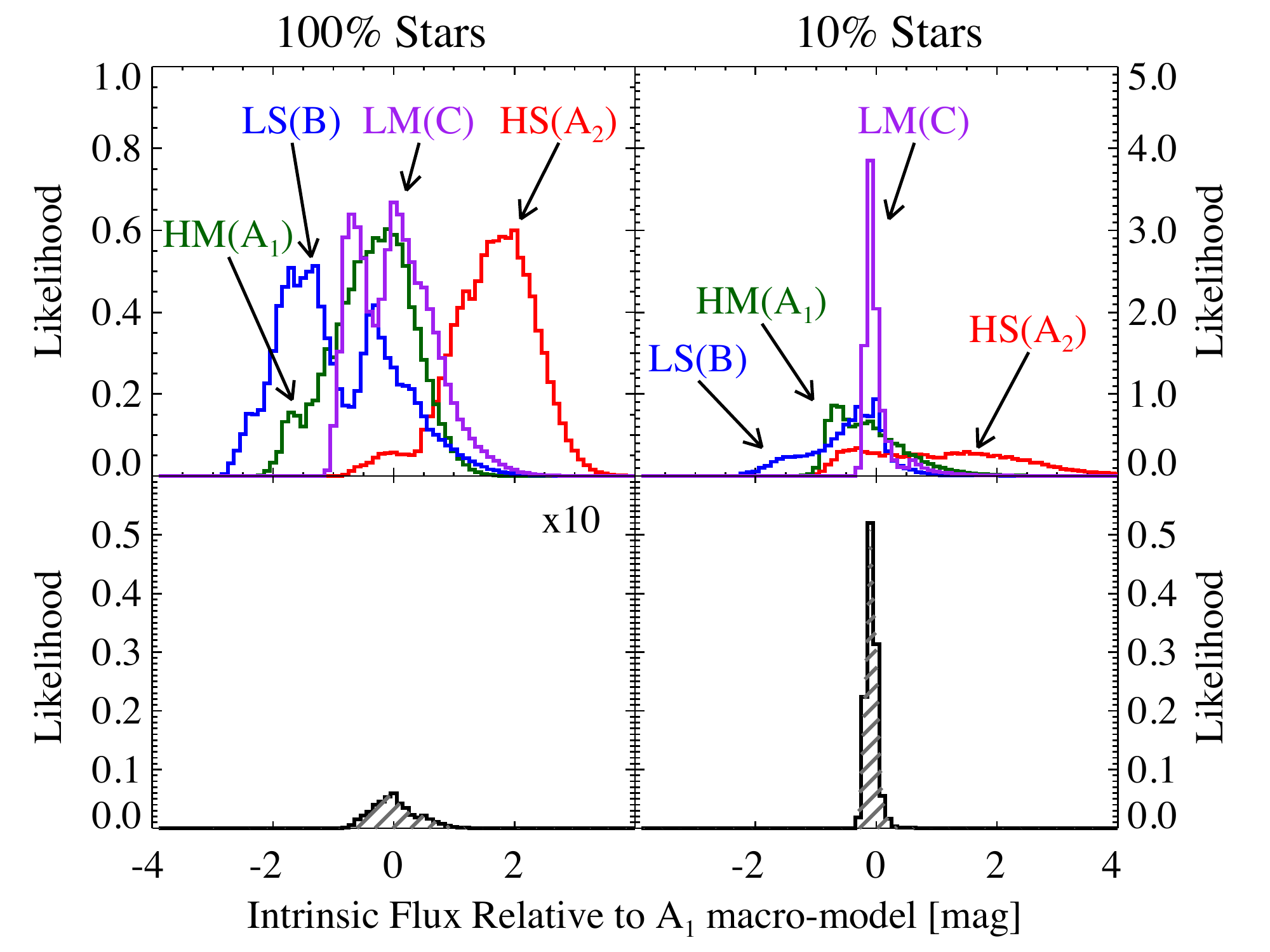}
\caption{Illustrative likelihood analysis for estimating the dark-to-stellar matter fraction.  Top panels: distributions expressing the likelihood of having an intrinsic X-ray source intensity, given the observed X-ray intensity.  The different colors are for the HS, HM, LS, and LM quad images (see text). The intensities are given in magnitude units.  The left panel is for a model where 100\% of the matter is in the form of stars; the right panel is for the case of 90\% dark matter and 10\% stars.  The histograms have each been shifted so that the zero point corresponds to the observed intensity, normalized to the smooth lens model value.  Bottom panels: distributions expressing the likelihood of having an intrinsic X-ray source intensity after taking into account the observed intensity of all four images combined.  Note that the histogram on the left (for 100\% stars) has been multiplied by a factor of 10 for ease in visibility. See text and Fig.\,\ref{fig:darkhist} for the results for other dark matter-to-star ratios. }
\label{fig:magdist}
\vglue0.5cm
\end{figure}
\vspace{0cm}

\subsection{Bayesian analysis 1}
The distributions of magnifications produced from such maps, for two different values of dark-to-stellar matter ratios, are shown in the top panels of Fig.\,\ref{fig:magdist} (the histograms have been shifted, which we describe below).  The different colored histograms in each panel are for the HS, HM, LS, and LM images.  These histograms represent $\mathcal{P}(O|I)$, which is the probability that if the source (the quasar) has an intrinsic intensity (i.e., in the absence of microlensing) $I$, an intensity $O$ will be observed (as modified by microlensing).  By Bayes' Theorem, this posterior probability is proportional to the likelihood that the intrinsic intensity is $I$ if we observe intensity $O$.

We treat the histograms in Fig.\,\ref{fig:magdist} as likelihood distributions of the intrinsic source intensity, given the observed fluxes of the four individual images (during the 2000 observation). Without loss of generality, we take the zeropoint to be the intrinsic source intensity such that the observed flux for image $A_1$ is exactly as predicted by the macro-model for \pg.  The histograms for the other images have been shifted by $[(m^\mathrm{obs}_X - m^\mathrm{obs}_{A_1}) - (m^\mathrm{macro}_X - m^\mathrm{macro}_{A_1})]$ to account for the fact that the observed magnitude differences do not agree with the macro-model magnitude differences and therefore must be micro-lensed. The largest shift is for the HS image (i.e., $A_2$ which was highly demagnified in 2000).  We then find the combined likelihood of an intrinsic intensity $I$ for all four images by taking the product of the (shifted) set of histograms. The results are shown in the bottom panels of Fig.\,\ref{fig:magdist} as likelihood distributions as a function of the intrinsic intensity of the source.  The area under the likelihood curve for the model with 100\% stars is an order of magnitude smaller than for the model with only 10\% stars (i.e., 90\% dark matter).

\begin{figure}[t]
\centering
\includegraphics[width=0.49\textwidth]{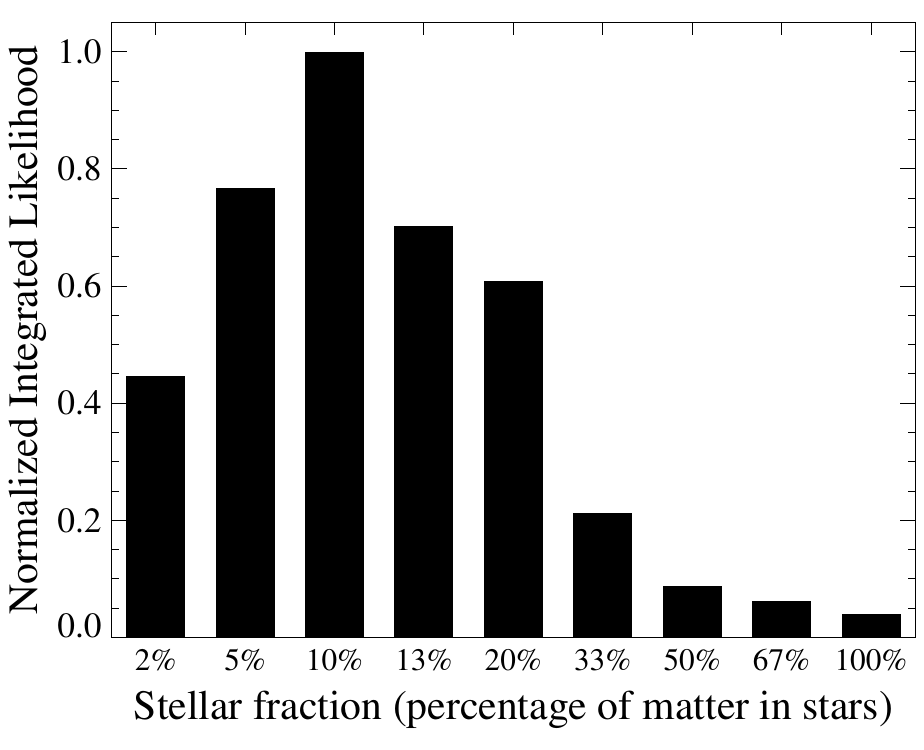}
\caption{Likelihood histogram of the dark-matter fraction based on the first maximum likelihood analysis described in the text.}
\label{fig:darkhist}
\end{figure}

We have repeated the same calculations for nine different stellar fractions.  The results are shown in Fig.\,\ref{fig:darkhist} as relative likelihood plotted against stellar fraction.  Note that the scale of stellar fraction is not linear.  From this result we conclude that the most likely fraction of mass in stars in the lensing galaxy at the typical impact parameter of the four lensed images ($\sim$5 kpc) is $\sim$10\%.

\subsection{Bayesian analysis 2}
We take a slightly different approach here to mitigating our ignorance of the intrinsic intensity of the source.  We perform an analysis based on each image's fraction of the total observed intensity (which is based on our two-dimensional fitting described in \S2.2).  The total {\em model} intensity is obtained by multiplying the microlensing map for each image by its macro-magnification and adding the resultant four maps together.  We then form fractional intensity maps by dividing the individual microlensing maps by the total intensity map.  

We use Bayes's theorem to calculate
\begin{eqnarray}
&& P(\mathrm{stellar\ fract}_i | f_{j,2000})  ~~ =  \nonumber \\
&&\hspace{1.0cm} \frac{P(f_{j,2000} | \mathrm{stellar\ fract}_i ) ~~ P(\mathrm{stellar\ fract}_i)}{P(f_{j,2000})} 
\label{eq:bayes2}
\end{eqnarray} 
where $f_{j,2000}$ is the observed flux (expressed as a fraction of the total intensity) of image $j$ $ \in (A_1, A_2, B, C$) in the 2000 observation and $\mathrm{stellar\ fract}_i$ is the stellar fraction for which a particular microlensing map (and its associated fractional intensity map) is generated.  We discuss each of the three terms on the right-hand side.

The term $P(f_{j,2000} | \mathrm{stellar\ fract}_i)$ represents the probability to obtain the observed intensity in 2000 for a specific stellar fraction.  To calculate this, we count all pixels in the fractional intensity map for that stellar fraction which lie within the 1-$\sigma$ confidence interval of $f_{j,2000}$ (e.g., $f_{A_2,2000} = 0.10\pm0.02$) and divide by the total number of pixels in the map.  

The term $P(\mathrm{stellar\ fract}_i)$ is the prior probability on a specific stellar fraction, which we take to be uniform.  All of these values are therefore $1/9$.

The denominator, $P(f_{j,2000})$, is similar to the first term but without regard to any particular stellar fraction.  We calculate this by counting the number of pixels in all maps that lie within the 1-$\sigma$ confidence interval of $f_{j,2000}$ and dividing by the total number of pixels in all maps.

\begin{figure*}[t]
\centering
\includegraphics[width=\textwidth]{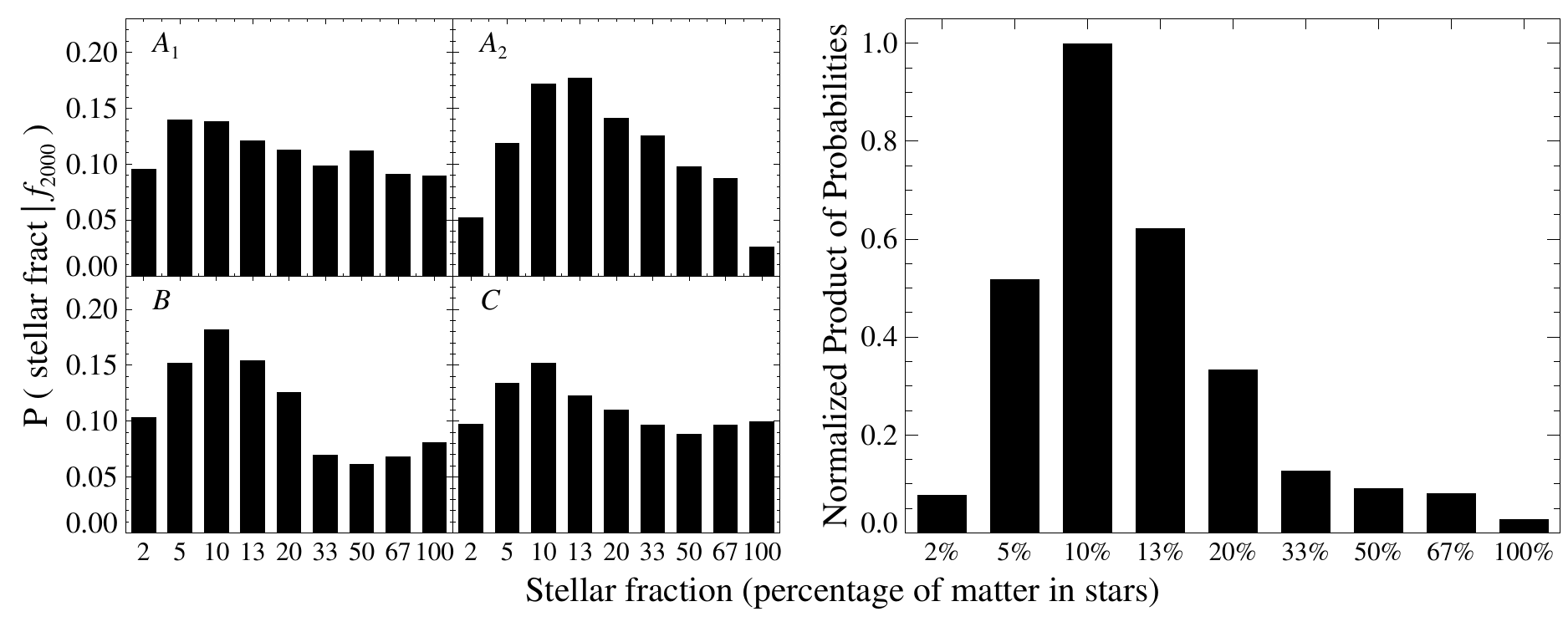}
\caption{{\it Left:} The probability of a certain stellar fraction given the observed flux in 2000 for each image, calculated using the second method described in the text.  {\it Right:} Product of these four probabilities for each stellar fraction, normalized to the highest value.}
\label{fig:darkhist2_2000}
\end{figure*}

Using this framework, we calculate the posterior probability of each stellar fraction based on each value of $f_j$.  We then multiply those probabilities for each stellar fraction together to produce a plot similar to Fig.~\ref{fig:darkhist}.  These results are shown in Fig.~\ref{fig:darkhist2_2000}.  The probabilities based on the individual images (left panels) show that some of the images have more power to discriminate between the different stellar fractions than others.  As expected, the $A_2$ image strongly favors certain fractions over others, and the $B$ and (to a lesser extent) $C$ images also show some power of discrimination.  The $A_1$ image appears roughly equally consistent with any stellar fraction.  Combining these individual image analyses, we obtain the right panel of Fig.~\ref{fig:darkhist2_2000}.

Because the previous analysis and this one utilize similar methods, one might expect them to yield identical answers.  Indeed, the right panel of Fig.~\ref{fig:darkhist2_2000} shows good agreement with the first Bayesian approach, namely, the most likely breakdown of matter in the lensing galaxy at the typical impact parameter of the four images is 10\% of the matter in stars and 90\% in a smooth, dark component.  The small differences between the two approaches are attributed to the fact that in the second approach, the position of the source is the same in all four maps, while in the first approach, the source will in general have different positions in the four different maps.

\subsection{Bayesian analysis 3}

The third approach utilizes two epochs of \chandra\ data.  We calculate the posterior probability of each stellar fraction given the observations in 2000 and 2008, subject to the constraint that the source and magnification map moved a certain amount relative to each other in the intervening eight years.  We assume a velocity range of 250--350~\kms\ in the lens plane. At the distance of the lens ($z_l = 0.31$), this corresponds to an angular movement of 0.33--0.46 $\mu$as over a time of $2800 / (1 + z_l)$ days.  For a standard $\Lambda$CDM cosmology the Einstein radius of a $0.7\Msun$ microlens in the lensing galaxy is $\theta_E \simeq 2.1 ~\mu$as.  Thus, the relative motion of the source and the map is 0.16--0.22 $\theta_E$ (see Fig.\,\ref{fig:Wambs}).  The distance would be nearly half an Einstein radius for a higher assumed speed of 1000~\kms\ (see Fig.\,\ref{fig:Wambs}).

We use Bayes's theorem to calculate
\begin{eqnarray}
&& P(\mathrm{stellar\ fract}_i | f_{j,2000} \bigcap f_{j,2008})  ~~ =  \nonumber \\
&& P(f_{j,2008} | \mathrm{stellar\ fract}_i \bigcap f_{j,2000}) ~~\frac{P(\mathrm{stellar\ fract}_i | f_{j,2000} ) }{P(f_{j,2008}|f_{j,2000})}~~~.
\label{eq:bayes3}
\end{eqnarray} 
We again discuss the three terms on the right-hand side.

The term $P(f_{j,2008} | \mathrm{stellar\ fract}_i \bigcap f_{j,2000})$ represents the probability to obtain the observed intensity in 2008 for a specific stellar fraction and observed intensity in 2000.  To calculate this, we first find all pixels, for a given stellar fraction, in the map which lie within the 1-$\sigma$ confidence interval of $f_{j,2000}$.  Around each of these pixels, we consider an annulus corresponding to our velocity range and the 8-year interval between observations;  the total number of pixels in all such annuli is $N_i$.  We also count the number $M_i$ of these pixels that lie within the 1-$\sigma$ confidence interval of $f_{j,2008}$, and we calculate $P(f_{j,2008} | \mathrm{stellar\ fraction}_i \bigcap f_{j,2000}) = M_i/N_i$.

The term $P(\mathrm{stellar\ fract}_i | f_{j,2000} )$ is the left-hand side of eq.~(\ref{eq:bayes2}) and was explained in the previous section.

The denominator, $P(f_{j,2008}|f_{j,2000})$, is similar to the first term but without regard to any particular stellar fraction.  We simply sum over all maps and calculate this term as $(\sum_i M_i) / (\sum_i N_i)$.

\begin{figure*}[t]
\centering
\includegraphics[width=\textwidth]{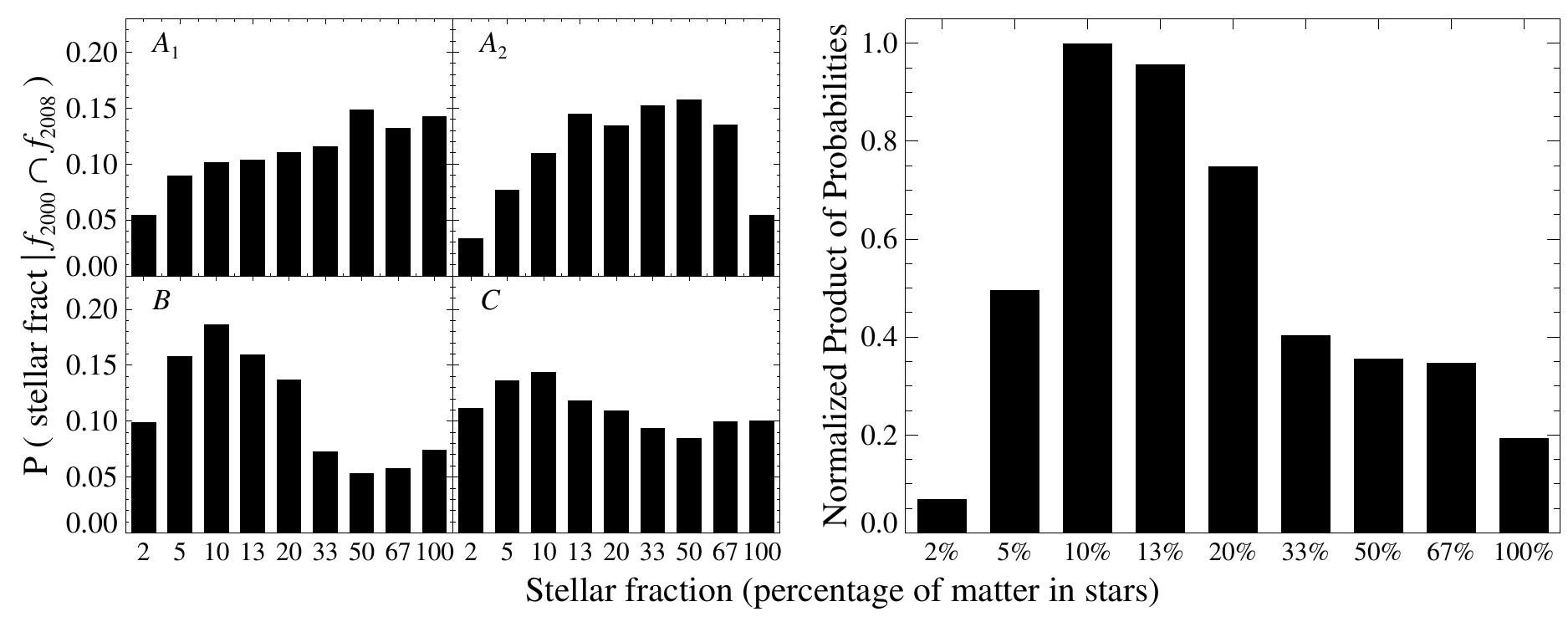}
\caption{{\it Left:} The probability of a certain stellar fraction given the observed fluxes in 2000 and 2008 for each image, as calculated using the third method described in the text.  {\it Right:} Product of the four probabilities for each stellar fraction, normalized to the highest value.}
\label{fig:darkhist2}
\end{figure*}

Similar to the previous analysis, we calculate the posterior probability of each stellar fraction based on each $f_j$.  We then multiply those probabilities for each stellar fraction together to produce a plot similar to Fig.~\ref{fig:darkhist2_2000}, and this is shown in Fig.~\ref{fig:darkhist2}.   Comparing this with the previous Bayesian analysis of just the first epoch of data, the most likely stellar fraction (10\%) is the same.  The addition of the 2008 data appears to have made smaller stellar fractions slightly less likely and larger stellar fractions slightly more likely.  We discuss this below.

A more comprehensive approach to multiply sampled lightcurves has been developed by \citet{2004ApJ...605...58K} who then applied it to the case of Q\,2237+0305.   The locations of the images of this quasar are close to the core of the (barred spiral) lensing galaxy because the lens redshift is low. This makes it much more likely that the quasar has crossed one or more caustics, but it also ensures a high stellar mass fraction, as was found by Kochanek.

\subsection{Evaluation of rapid temporal change}
We use the stellar microlensing magnification maps to investigate the likelihood that the $A_2$ image would have changed its intensity by a factor of $\sim$6 during an interval of eight years (in the observer frame). We choose this interval because it is the length of time between the first \chandra\ observation of \pg\ and the most recent.  \xmm\ observations during this interval indicate that the rise in $A_2$ could have occurred as early as November 2001 \citep[see Fig.~3 of][]{2006ApJ...648...67P}, but the \chandra\ observation in 2008 is the first to show directly that $A_2$ has brightened.  In Fig.\,\ref{fig:bayes2} we plot the probability of finding image $A_2$ with a fraction $f_{A_2,2008}$ of the total source intensity in 2008, given that the fraction in 2000 was $f_{A_2,2000} = 0.10\pm 0.02$.  This probability density function is shown for each of the nine stellar fractions that we considered.  These functions are computed as described above for $P(f_{j,2008} | \mathrm{stellar\ frac}_i \bigcap f_{j,2000})$ with $j=A_2$ for a complete range of possible values for $f_{A_2,2008}$.  As can be seen, it is unlikely in all cases to observe such a large rise in $A_2$ in only eight years.

\subsection{Discussion of the Bayesian methods}
The individual panels of the left-hand side of Fig.~\ref{fig:darkhist2_2000} are essentially another way of looking at the magnification map histograms in the top panels of Fig.~\ref{fig:magdist}; the panels of Fig.~\ref{fig:darkhist2_2000} show the probabilities of the nine magnification maps to produce the flux fraction that was observed.  Whereas all maps (i.e., stellar fractions) can accommodate the $A_1$ and, to some extent, $C$ observed values, the $A_2$ and $B$ values observed in 2000 are much more likely to have come from stellar fractions of 5--20\%, producing a combined probability that is fairly strongly peaked around a stellar fraction of 10\%.  

When the additional information that $A_2$ became much less demagnified on a timescale of eight years is added, the combined probability becomes less strongly peaked.  The individual panels on the left-hand side of Fig.~\ref{fig:darkhist2} show that, in comparison to the 2000 data alone, this difference is due mainly to the difference in the $A_2$ probability distribution, which in this case is roughly equally probable to come from a stellar fraction of 10\% as 67\%.  We interpret this as a reflection of the relatively higher probability to cross a caustic on such a short timescale in high-stellar-fraction magnification maps; in low-stellar-fraction magnification maps, the caustics are too few and far between.  Put another way, there is a tradeoff between having a low enough stellar density to give a high probability to find $A_2$ in a demagnified state (in 2000) and having a high enough stellar density to have enough caustics that $A_2$ can become brightened on a short time scale (in 2008). For example, Fig.~\ref{fig:bayes2} shows that, although all stellar fractions give a low probability of observing such a dramatic change in $A_2$ in only eight years, the highest stellar fractions give a higher probability of such a change.

If the relative motion between the source and the magnification map were much larger, it would be easier to accommodate both the 2000 and 2008 observations of $A_2$ with the low-stellar-fraction maps, as it would be if the temporal baseline were much larger, which it very well could be.  Fig.~3 of \citet{2006ApJ...648...67P} shows that, based on the unresolved total flux, it is likely that $A_2$ was demagnified in all previous X-ray observations except the first observation by {\it Einstein} in December 1979 which showed a total flux from \pg\ comparable to that of the first \xmm\ observation in November 2001.  Although the X-ray light curve is sparsely sampled, and although \chandra\ is the only instrument which can separately measure the individual images, it is possible that $A_2$ was in a demagnified state for $\sim$22 years.  This would certainly change the results of the second Bayesian analysis we performed.

In addition to the uncertainties in baselines (due to sparsely sampled light curves) and relative velocities, there is another shortcoming to this type of Bayesian analysis, namely, the use of static magnification maps to analyze temporal behavior.  As shown by \citet{1993ApJ...404..455K, 1995ApJ...450...19W}, the motions of the individual stars in the lensing galaxy can cause shorter and more frequent microlensing events than bulk motion can produce, by perhaps a factor of two. Unlike the present case, the magnification maps they used had zero shear.  A preliminary reconaissance of maps with shear equal to the convergence, appropriate to isothermal potentials, and with 10-20\% of the mass in stars, indicates that allowing for the motion of individual stars would again produce changes on a timescale only a factor of two shorter.

Given the additional complications introduced by the motions of the quasar and the microlensing stars, the first and second Bayesian analyses are clearly more straightforward, free from the uncertainties in the third.  In a forthcoming paper, we will apply the first method to 14 quadruply lensed quasars that \chandra\ has observed (Pooley et al.\ in preparation).

Finally, we point out one common feature of all three approaches: the important role that the low-magnification images ($B$ and $C$) play in constraining the stellar fraction.  This is evident in the individual panels of Figs.~\ref{fig:darkhist2} and \ref{fig:darkhist2_2000}.  It can also be seen in the upper panels of Fig.~\ref{fig:magdist} by considering the effects of the LM and LS histograms (their peaks and their cutoffs) on the final product.

\section{Summary and Conclusions}
We have observed a dramatic change in the X-ray flux of the $A_2$ image of \pg\ in \chandra\ observations that were separated by eight years.  The short timescale for the flux change clearly indicates the presence of stellar microlensing rather than milli-lensing due to dark matter substructure in the elliptical lensing galaxy.  The observations of the individual fluxes point toward a substantial dark matter fraction of $\sim 80-95$\%.  

\begin{figure}[t]
\centering
\includegraphics[width=0.49\textwidth]{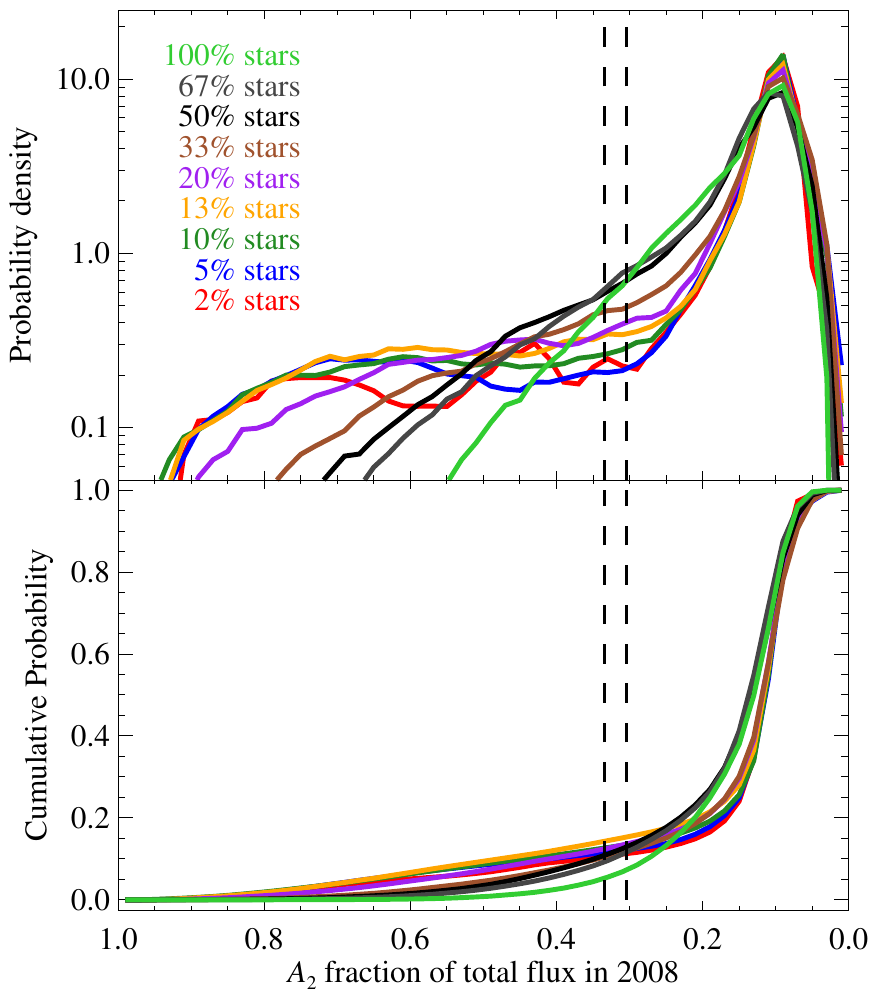}
\caption{Probability density ({\it top}) and cumulative probability ({\it bottom}) of $A_2$ fractional flux in 2008 given the value in 2000.  The dashed lines mark the 1-$\sigma$ confidence intervals of the measured value in 2008.}
\label{fig:bayes2}
\end{figure}

One particularly interesting aspect of the observed change in flux ratio between the $A_2$ and $A_1$ images is the very rapid timescale on which it occurred ($\sim$8 years).  Figure \ref{fig:Wambs} indicates how far bulk translation of the lensing galaxy is likely to move the quasar image with respect to the caustic pattern during an eight-year interval.  For typical expected transverse speeds of $\sim$300 km s$^{-1}$ it seems unlikely that the image will start on a point of very low magnification and end up with near nominal magnification in just over eight years, as indicated in Fig.~\ref{fig:bayes2}.  With an unexpectedly high speed of $\sim$1000 km s$^{-1}$ the likelihood for the observed change in magnification of the $A_2$ image becomes considerably greater.  In this regard, we note that the recent report of an even larger change in a flux ratio in the quad lens RX\, J1131$-$1231 over an interval of $\sim$2 years \citep{Chartas} indicates that the variation that we observed in \pg\ is perhaps not a rare occurrence.  The difference between the effect of bulk and random motion of the microlensing stars is probably not sufficient to account for these rapid variations, leaving us with an unresolved puzzle.

\acknowledgements 

D.~P.\ thanks Nicholas E.\ Matsakis and Eric M.\ Downes for extremely useful discussions of Bayesian analysis and is grateful to the MIT Kavli Institute for Astrophysics and Space Research for its hospitality during the summer of 2008 during which most of this work was done.  D.~P.\ and S.~R.\ gratefully acknowledge support from \chandra\ grants G07-8098A\&B.   J.~A.~B.\ and P.~L.~S.\ acknowledge support from US NSF grant AST-0602010.

\end{document}